\documentclass[usegraphicx,usenatbib]{mn2e}

\begin{document}
\title[Radio and mid-infrared correlations]{A New Radio Loudness Diagnostic for Active Galaxies: a Radio-To-Mid-Infrared Parameter}
\author[Mel\'endez et. al]{M. Mel\'endez$^{1,2}$\thanks{E-mail:marcio.b.melendez@nasa.gov}, S.B. Kraemer$^{1,3}$, H.R. Schmitt$^{4,5}$\\
$^1$Astrophysics Science Division, NASA Goddard Space Flight Center, Greenbelt, MD, 20771, USA\\
$^2$Department of Physics and Astronomy, Johns Hopkins University, Baltimore, MD, 21218, USA\\
$^3$Institute for Astrophysics and Computational
Sciences, Department of Physics, The Catholic University of America,
 Washington, DC, 20064, USA\\
$^4$Remote Sensing Division, Naval Research Laboratory, Washington, DC, 20375,  USA\\
$^5$Interferometrics, Inc., Herndon, VA, 20171, USA
}

\maketitle

\begin{abstract}
We have studied the relationship between the nuclear (high-resolution) radio emission, at 8.4~GHz  (3.6~cm) and 1.4~GHz (20~cm), the [O~IV]~$\lambda$25.89~$\micron$, [Ne~III]~$\lambda$15.56~$\micron$  and  [Ne~II]~$\lambda$12.81~$\micron$ emission lines and the black hole mass accretion rate for a sample of Seyfert galaxies. In order to characterize the radio contribution for the Seyfert nuclei we used the 8.4GHz/[O~IV] ratio, assuming that [O~IV] scales with the luminosity of the AGN. From this we find that Seyfert 1's (i.e., Seyfert 1.0's, 1.2's, and 1.5's)  and Seyfert 2's (i.e., Seyfert 1.8's, 1.9's, and 2.0's) have similar radio contributions, relative to the AGN. On the other hand, sources in which the [Ne~II] emission is dominated either by the AGN or  star formation  have statistically different radio contributions, with star formation dominated sources more ``radio loud", by a factor of $\sim 2.8$ on average, than AGN dominated sources. We show that star formation dominated sources with  relatively larger  radio contribution have smaller mass accretion rates. Overall, we suggest that  8.4GHz/[O~IV], or alternatively, 1.4GHz/[O~IV] ratios, can be used to characterize the radio contribution, relative to the AGN, without the limitation of previous methods that rely on optical observables. 

\end{abstract}


\section{Introduction}
Active galactic nuclei (AGN) are among the most powerful objects in the universe with an extraordinary amount of energy released from their  unresolved nuclei.  Active super-massive black holes (SMBH) in the center of these galaxies are responsible for the tremendous diversity in their observed energy emission \citep[e.g.,][]{1984ARA&A..22..471R,1989ApJ...347...29S,2000ApJ...540L..13P,2004ApJ...613..682P}. Despite the fact that the unified  scheme of AGN has been able to explain the observed type I/II dichotomy \citep[see][for a review]{1993ARA&A..31..473A}, it is clear that by looking at different wave bands there are additional parameters that can complement our current understanding of AGN classification other than to be attributed solely to a viewing angle difference \citep[e.g.,][]{1997ApJ...485..552M,2001ApJ...546..845G,2006AJ....132..401B,2008ApJ...689...95M}. Therefore, in order to unveil and understand the complexity of these sources, ones needs to rely  on multiwavelength studies. In particular,  the radio  and mid-infrared emissions are basically unaffected by  dust extinction which make these energy bands ideal to investigate the nature of the ``hidden" nuclear engine. On the other hand, extended radio emission and low-ionization mid-infrared emission lines  are excellent tracers of star formation activity when properly calibrated \citep[e.g.,][]{1992ARA&A..30..575C,2007ApJ...658..314H}.

 Furthermore,  the strength of the observed radio emission in AGN seems to follow a bimodal distribution falling into  ``radio loud" and ``radio quiet" sources \citep[e.g.,][]{1990MNRAS.244..207M,1999AJ....118.1169X}. Historically, the dividing criterion for the AGN radio loudness was adopted to be the ratio between the radio emission, at the most commonly used frequency of 5~GHz, and the optical continuum at $\sim 4400$~\AA, i.e., the  radio-to-optical parameter, ${\rm R=L_\nu{\rm (6{\rm cm})}/L_\nu({\rm B})}$ \citep[e.g.,][]{1989AJ.....98.1195K,1992ApJ...391..560V}. This convention to quantize the radio loudness of a source is straightforward  when the contribution from the host galaxy does not affect the observed radio and optical components of the AGN, for example in quasars, where the active nucleus overpowers the underlying galaxy. In the case of Seyfert galaxies the quantification  of their radio loudness could be  more subtle.  \citet{2001ApJ...555..650H} found that most Seyfert 1 galaxies, previously classified as radio-quiet, fall  into the radio loud regime when the nuclear component for the radio-to-optical parameter is properly measured. However, even with a resolution of few arcseconds, the nuclear  component of the optical luminosity could be contaminated by star formation and   is strongly affected by dust extinction that may lead to an erroneous AGN characterization.

Previous multiwavelength studies have shown  statistically significant correlations  between the radio core luminosity and purportedly orientation-independent measures of the intrinsic luminosity of the active nuclei, which   suggests a common mechanism at nuclear (subkiloparsec) scales, , e.g., accretion onto the black hole \citep[e.g.,][]{1989MNRAS.240..701R,1999AJ....118.1169X,2002ApJ...564..120H,2009ApJ...698..623D}. On the other hand, the correlations between extra-nuclear radio emission and the far-infrared (FIR) luminosity of the galaxy suggest a circumnuclear starburst as the common cause for these observables \citep[e.g.,][]{1982ApJ...252..102C,1988AJ.....96...30C,1993ApJ...419..553B,1998MNRAS.301.1019R}.  In this regard, early studies suggested two possible scenarios for the origin of the compact radio fluxes observed in radio quiet sources: either the radio emission is associated with a nuclear starburst or generated by relativistic electrons accelerated by a low-power jet (similar to radio loud sources), with high-resolution radio-imaging observations favoring the latter \citep[e.g.,][]{1998MNRAS.299..165B}. More recently, the good correlation between the X-ray and the radio luminosities  raise the possibility  that radio emission in radio quiet sources may  have  a coronal origin \citep{2008MNRAS.390..847L}. All these exemplify that the radio properties of the AGN are diverse and can be intimately  associated with fundamental parameters of the active galaxies, such as the black hole masses, mass accretion rates, and  host galaxy morphology \citep[e.g.,][]{1999AJ....118.1169X,2000ApJ...543L.111L,2001A&A...380...31W}.

Recently,  \citet{2008ApJ...682...94M} found a tight correlation in Seyfert 1 galaxies between the [O~IV]~$\lambda$25.89~$\micron$ and the X-ray 14-195~keV continuum luminosities from  {\it Swift}/BAT observations \citep{2005ApJ...633L..77M,2008ApJ...681..113T}. A weaker correlation was  found in Seyfert 2 galaxies, which was attributed to the effect of Compton scattering  in the 14-195~keV band. These results suggest  that [O~IV] is a truly isotropic property of AGNs, given its high ionization potential and that it is basically unaffected by reddening, meaning that the [O~IV] strength directly measures the AGN power. This result has been later confirmed in more  complete samples \citep[][Weaver et al. submitted to ApJ]{2009ApJ...700.1878R}. Moreover, the strong correlation found between [O~IV] and the high-ionization ($\sim$97~eV) [Ne~V]~$\lambda$14.32/24.32~$\micron$ emission is a further argument for [O~IV] being a good indicator for the intrinsic luminosity of the AGN \citep[e.g.,][Weaver et al. submitted to ApJ]{2009ApJ...691.1501D,2009MNRAS.398.1165G}. It is the primary goal of this paper to examine the physical relationship between the nuclear radio emission and the mid-infrared emission line luminosities in Seyfert galaxies. We use high-resolution radio continuum measurements and the [O~IV]  emission line to investigate  the radio loudness of the Seyfert nuclei  and the black hole mass accretion rate for a sample of Seyfert galaxies.

\section{The Sample}
\begin{figure}
\includegraphics[width=84mm]{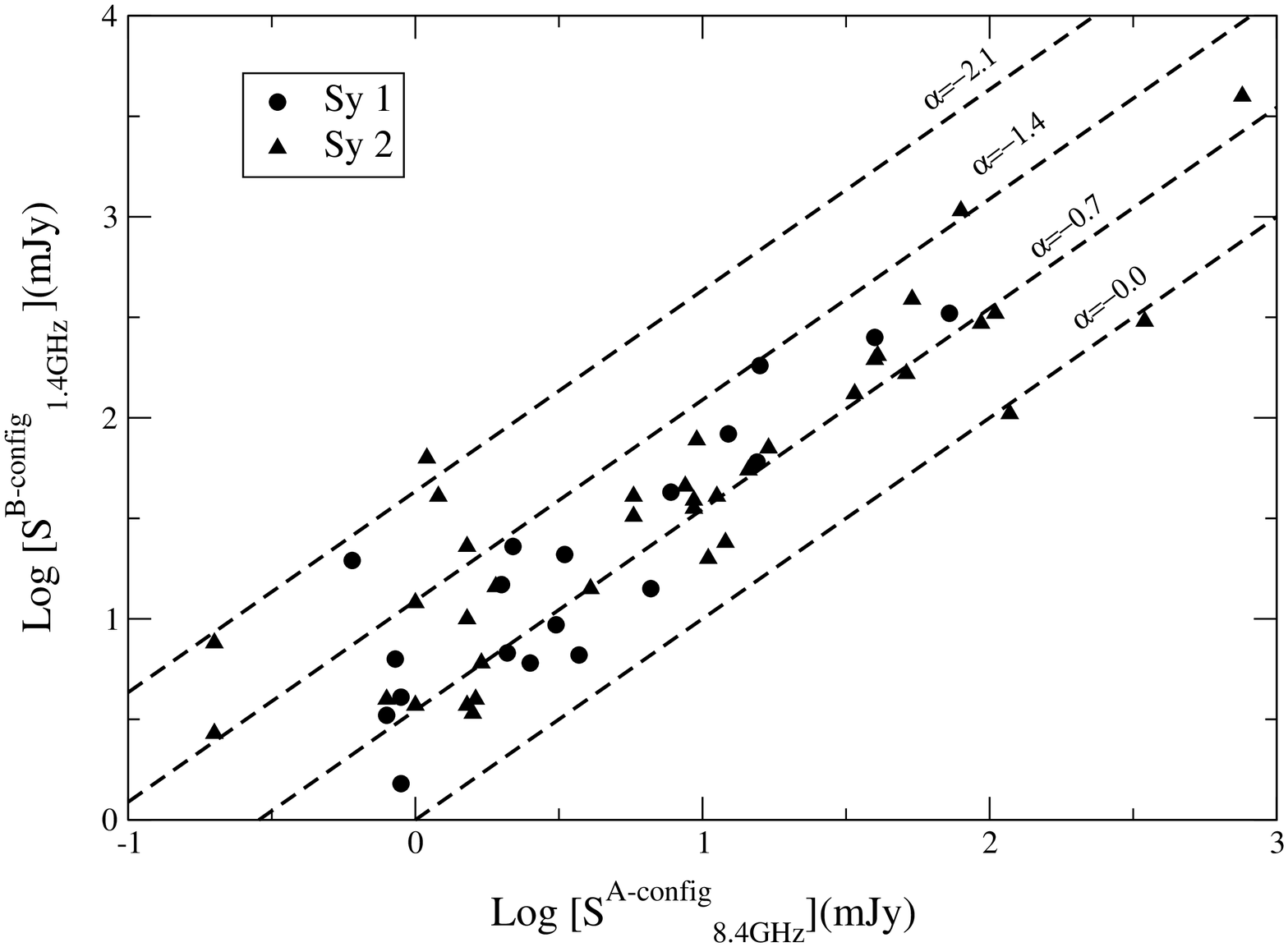}
    \caption{Correlation between the high-resolution 1.4~GHz and 8.4~GHz. The dashed lines indicate the flux density  relationship expected for different spectral indices in Seyfert~1 ({\it circles}) and Seyfert~2 ({\it triangles}) galaxies.\label{fig1}}
\end{figure}

We have obtained, from the literature, observations with the Very Large Array (VLA) in A-configuration at 8.4-GHz \citep[see][for instrumental details]{1980ApJS...44..151T}, with a resolution of $0.25$$'$$'$ and the largest detectable size of $3.5$$'$$'$, for  the radio quiet sources  in the sample of Seyfert galaxies  presented in \citet{2008ApJ...689...95M}. This sample of Seyfert galaxies includes  compilations from \citet{2007ApJ...671..124D}, \citet{2007arXiv0710.4448T}, \citet{2002A&A...393..821S},  \citet{2005ApJ...633..706W} and from our analysis of archival spectra observed with the Infrared Spectrograph (IRS) on board {\it Spitzer Space Telescope} \citep{2004ApJS..154....1W} (see Appendix). For comparison, we also have obtained from the literature  high resolution  1.4~GHz data  observed with the VLA in B configuration, which has a resolution of $\sim 5$$'$$'$, thus only encompassing the nucleus. Figure~\ref{fig1} shows the good correlation between the 1.4~GHz and 8.4~GHz fluxes and the distribution of their  spectral indices. The good correlation is expected as in high-resolution both fluxes are dominated by the active nucleus. One needs to consider that the spectral indices  may provide a lower limit to the nuclear component, as the  1.4~GHz B-configuration fluxes  could  include some extranuclear emission, thus a possible contamination from star formation. In this regard, and in order to  complete this set of high-resolution 8.4~GHz fluxes, we extrapolated high-resolution observations at different frequencies to 8.4GHz by assuming a conservative $f_\nu \propto \nu^{-0.7}$, as determined  by the spectral indices distribution. Finally, in order to check on the  effect of radio emission regions on scales bigger than that sampled by the VLA in B configuration at 1.4~GHz, we compared these data  with fluxes from the NRAO VLA Sky Survey (NVSS), which has a resolution of $\sim 45$$'$$'$. We found an excellent correlation between VLA in B configuration at 1.4~GHz and NVSS fluxes, $r_s=0.89,~P_r<10^{-6}$, that suggests that, in our sample, compact nuclear regions dominate over large-scale extranuclear emission regions.

The  final sample includes  29 Seyfert 1's (i.e., Seyfert 1.0's, 1.2's, and 1.5's)  and 59 Seyfert 2's (i.e., Seyfert 1.8's, 1.9's, and 2.0's), which are listed in Table~\ref{table1}.  There are  six upper limits for the 8.4~GHz emission, two Seyfert 1 galaxies and four  Seyfert 2 galaxies.  Note that it was not possible to find 8.4~GHz  high-resolution fluxes for all the galaxies in the sample first presented by  \citet{2008ApJ...689...95M}. The total radio luminosity, ${\rm L_{8.4GHz}\equiv \nu L_\nu {\rm (8.4GHz)}}$ (ergs/s), and the different mid-infrared emission-line luminosities are calculated from the fluxes assuming $H_o=71{\rm kms^{-1}Mpc^{-1}}$ and  a deceleration parameter $q_o=0$.  At the median redshift for the sample, $z= 0.02$, 1$'$$'$ represents $\sim$400~pc. Furthermore, for a full aperture extraction, {\it Spitzer} short-low (low-resolution, R$\sim 60-127$) and short-high (high-resolution, R$\sim 600$) slits will  sample, in the dispersion direction, a region equivalent to the region sampled by the VLA in B-configuration at 1.4-GHz.

\begin{table*}
 \centering
 \begin{minipage}{140mm}
\caption{The Infrared Sample of Seyfert Galaxies\label{table1}}
 \begin{tabular}{@{}lllccccccccl@{}}
  \hline
Name  &$z$&  Type& ${\rm L_{8.4GHz}}$& ${\rm L_{1.4GHz}}$
& $\log{\rm M_{BH}}$&$\log {\rm L_{bol}/L_{Edd}}$ &$ {\rm R_{[O IV]}}$ &  $ {\rm R_{[Ne III]}}$ &
  $ {\rm R_{[Ne II]}}$ &  MC &  Ref. \\

&&& \multicolumn{2}{c}{($\log$~ergs ${\rm s^{-1}}$)}&$(M/M_\odot)$&\\

 \hline

CGCG381-051&0.0307&2&38.02& & & &-2.30&-2.77&-3.53&S &2\\
Circinus&0.0014&2&37.10$^{a}$& &6.23$^{b}$&-0.19&-3.38&-3.08&-3.51&S &8,24\\
ESO012-G021&0.0300&1&$<$39.08$^{a}$& & & &-2.42&-2.03&-2.30&A&9\\
ESO103-G035&0.0133&1&38.79$^{a}$& &7.15&-0.47&-2.34&-2.20&-2.09&A&10,25\\
ESO141-G055&0.0360&1&$<$39.24$^{a}$& &7.10$^{b}$&-0.23&-2.08&-1.97&-1.57&A&9,26\\
ESO541-G012&0.0566&2&38.69& & & &-2.87&-2.48&-2.44&A&2\\
ESO545-G013&0.0337&1&38.65& & & &-2.82&-2.77&-2.76&S &2\\
I Zw 1&0.0611&1&38.78&38.88&7.26$^{d}$&-0.15&-2.77&-3.24&-2.96&S &3,14,27\\
IC4329A&0.0161&1.2&38.86&38.67&7.00$^{e}$&0.32&-2.91&-2.55&-2.20&A&2,13,28\\
IRAS00198-7926&0.0728&2&40.59$^{a}$& &8.44$^{b}$&-0.28&-2.01&-1.64&-1.28&A&11,29\\
IRAS00521-7054&0.0689&2&40.64$^{a}$& &8.45$^{b}$&-0.92&-1.33&-1.30&-1.16&S &11,30\\
IRAS01475-0740&0.0177&2&39.89& &6.09&0.14&-0.79&-1.02&-1.15&S &1,31\\
IRAS04385-0828&0.0151&2&38.50& &8.77$^{b}$&-2.44&-2.27&-2.45&-2.58&S &1,32\\
IRAS15091-2107&0.0446&1&39.46&39.81&7.00$^{d}$&0.69&-2.68&-2.40&-2.25&A&2,16,33\\
IRAS15480-0344&0.0303&2&39.28&39.06&8.22$^{b}$&-0.83&-2.56&-2.20&-1.87&A&1,15,29\\
IRAS22017+0319&0.0611&2&39.15& &8.07$^{b}$&-0.13&-3.24&-2.92&-2.55&A&2,30\\
MCG-2-40-4&0.0252&2&38.54& & & &-2.85&-2.85&-2.74&A&2\\
MCG-2-58-22&0.0469&1.5&39.74$^{a}$& &7.18$^{b}$&0.08&-1.96&-1.95&-1.54&A&4,26\\
MCG-2-8-39&0.0299&2&38.52& &7.85&-0.82&-2.95&-2.77&-2.62&A&1,31\\
MCG-3-58-7&0.0315&2&38.22& & & &-3.20&-3.12&-2.82&A&2\\
MCG-5-13-17&0.0126&1.5&37.86&37.46&7.94&-1.77&-2.76&-2.79&-2.72&A&1,13,31\\
MCG-6-30-15&0.0077&1.2&36.99&36.86&6.81&-0.78&-3.49&-2.74&-2.94&A&13,13,31\\
MRK 3 &0.0140&2&39.45&39.80&8.36&-0.86&-2.50&-2.49&-2.11&A&13,13,31\\
MRK 6&0.0188&1.5&39.41&39.44&7.64$^{b}$&-0.50&-2.18&-2.15&-1.86&A&1,13,26\\
MRK 9&0.0399&1.5&38.38&38.21&7.46$^{b}$&-0.62&-2.92&-2.45&-2.68&A&2,15,26\\
MRK 79&0.0222&1.2&38.26&38.35&7.72$^{e}$&-0.43&-3.48&-3.08&-2.77&A&1,15,28\\
MRK 273&0.0378&2&39.95&39.77&8.22&-0.37&-2.36&-2.18&-2.28&S &2,15,31\\
MRK 334&0.0219&1.8&38.12&38.53&6.51$^{b}$&0.24&-3.08&-3.31&-3.38&S &3,12,34\\
MRK 335 &0.0258&1.2&38.41&38.14&7.15$^{e}$&-0.57&-2.62&-2.13&-2.07&A&3,13,28\\
MRK 348&0.0150&2&40.15&39.32&6.83&-0.21&-0.92&-0.93&-0.65&A&2,13,31\\
MRK 471&0.0342&1.8&38.55$^{a}$&38.16&7.74 &-1.17&-2.47&-2.34&-2.34&A&12,12,35\\
MRK 509&0.0344&1.2&39.17$^{a}$&38.72&8.16$^{e}$&-0.86&-2.58&-2.47&-2.37&A&4,17,28\\
MRK 573&0.0172&2&38.01&38.11&7.28&-0.02&-3.69&-3.18&-2.91&A&3,13,36\\
MRK 609&0.0345&1.8&39.20$^{a}$& &6.80$^{f}$&0.08&-2.13&-2.18&-2.30&S &4,37\\
MRK 622&0.0232&2&38.23&38.00&6.92&-0.30&-2.85&-2.75&-2.62&A&13,13,36\\
MRK 817&0.0315&1.5&38.82& &7.69$^{e}$&-1.02&-2.30&-2.12&-2.22&A&3,28\\
MRK 883&0.0375&1.9&39.43$^{a}$&38.94&8.15&-1.10&-2.06&-1.96&-2.10&S &12,12,25\\
MRK 897&0.0263&2&38.65& &7.42 &-1.89&-1.32&-2.17&-2.91&S &2,25\\
NGC 424&0.0118&2&38.48&38.00&7.54 &-1.26&-2.26&-2.32&-2.18&A&2,13,25\\
NGC 513&0.0195&2&37.92&38.68&7.65&-1.22&-2.95&-2.90&-3.04&S &2,13,36\\
NGC 931&0.0167&1.5&37.78& &7.64 &-0.67&-3.63&-3.34&-2.90&A&1,36\\
NGC 985&0.0431&1&39.11$^{a}$&38.58&8.05$^{d}$&-0.76&-2.63&-2.58&-2.36&A&4,15,38\\
NGC 1068&0.0038&2&39.30&39.24&7.56 &-0.24&-2.47&-2.40&-2.04&A&3,15,25\\
NGC 1125&0.0109&2&38.40& &7.01 &-0.58&-2.47&-2.50&-2.44&A&2,25\\
NGC 1194&0.0136&1&37.48&36.92&7.32$^{b}$&-1.02&-3.27&-3.08&-2.82&A&2,15,39\\
NGC 1320&0.0135&2&37.52&37.31&7.18&-0.52&-3.58&-3.03&-3.03&A&2,13,36\\
NGC 1365&0.0055&1.8&37.70&37.54&7.62$^{g}$&-1.03&-3.34&-2.89&-3.33&S &2,18,40\\
NGC 1667&0.0152&2&37.80&37.41&7.62 &-1.29&-2.98&-3.18&-3.31&S &2,19,25\\
NGC 2639&0.0111&1.9&39.42&38.59&7.94 &-2.48&-0.48&-0.78&-1.05&S &2,19,41\\
NGC 2992&0.0077&2&38.64&38.57&7.87 &-1.19&-2.48&-2.26&-2.21&A&2,20,25\\
NGC 3079&0.0037&2&38.37&38.09&7.97 &-2.39&-1.66&-1.55&-2.28&S &3,15,41\\
NGC 3227&0.0039&1.5&37.51&37.57&7.63$^{e}$&-1.78&-2.77&-2.85&-2.80&A&3,15,28\\
NGC 3516&0.0088&1.5&37.64&37.35&7.63$^{e}$&-1.27&-3.16&-3.00&-2.49&A&3,13,28\\
NGC 3660&0.0123&2&$<$36.92& &6.83 &-1.21&-3.16&-2.77&-3.41&S &2,25\\
NGC 3783&0.0097&1.5&38.14& &7.47$^{e}$&-1.14&-2.65&-2.59&-2.49&A&1,28\\
NGC 3786&0.0089&1.8&37.16&37.46&7.53&-1.46&-3.35&-3.15&-3.12&A&3,20,36\\
NGC 3982&0.0037&2&36.29&36.22&6.37 &-2.05&-2.47&-2.87&-3.38&S &3,15,41\\

\hline
\end{tabular}
\end{minipage}
\end{table*}

\begin{table*}
 \centering
 \begin{minipage}{140mm}
\contcaption{The Infrared Sample of Seyfert Galaxies}
 \begin{tabular}{@{}lllccccccccl@{}}
  \hline
Name  &$z$&  Type& ${\rm L_{8.4GHz}}$& ${\rm L_{1.4GHz}}$
& $\log{\rm M_{BH}}$&$\log {\rm L_{bol}/L_{Edd}}$ &$ {\rm R_{[O IV]}}$ &  $ {\rm R_{[Ne III]}}$ &
  $ {\rm R_{[Ne II]}}$ &  MC &  Ref. \\

&&& \multicolumn{2}{c}{($\log$~ergs ${\rm s^{-1}}$)}&$(M/M_\odot)$&\\

 \hline

NGC 4051&0.0023&1.5&35.77&36.50&6.28$^{e}$&-1.16&-3.80&-3.66&-3.53&A&3,15,28\\
NGC 4151&0.0033&1.5&38.16&38.04&7.12$^{e}$&-0.81&-2.61&-2.61&-2.35&A&3,15,28\\
NGC 4388&0.0084&2&38.08&37.88&7.10 &0.04&-3.50&-3.09&-3.03&A&3,15,25\\
NGC 4501&0.0076&2&$<$36.32&36.67&7.81 &-2.54&-3.40&-3.45&-3.62&S &2,15,41\\
NGC 4507&0.0118&1.9&38.11& &7.56 &-0.94&-2.95&-2.83&-2.90&A&5,25\\
NGC 4748&0.0146&1&37.70& &6.44&0.69&-3.88&-3.16&-2.83&A&5,31\\
NGC 4941&0.0037&2&37.00&36.77&6.88 &-1.46&-2.86&-2.84&-2.63&A&1,19,25\\
NGC 4968&0.0099&2&38.01&37.98&7.25 &-0.90&-2.79&-2.69&-2.69&A&2,13,25\\
NGC 5005&0.0032&2&37.20&37.14&7.87 &-2.73&-2.39&-2.34&-2.91&S &2,15,41\\
NGC 5256&0.0279&2&39.32&39.12&6.92$^{b}$&0.65&-2.70&-2.48&-2.80&S &2,15,26\\
NGC 5347&0.0078&2&37.24&36.79&6.79&-1.30&-2.69&-2.49&-2.59&A&1,19,36\\
NGC 5506&0.0062&1.9&38.86&38.58&6.88 &-0.12&-2.35&-1.93&-1.77&A&1,15,25\\
NGC 5548&0.0172&1.5&38.07&38.31&7.83$^{e}$&-1.29&-2.91&-2.88&-2.34&A&3,13,28\\
NGC 5643&0.0040&2&37.22&37.29&6.79 &-0.64&-3.38&-3.06&-2.97&A&6,21,25\\
NGC 5929&0.0083&2&38.32&38.17&7.22&-1.65&-1.70&-1.80&-2.17&S &3,15,31\\
NGC 5953&0.0066&2&36.93&37.92&6.79 &-0.93&-3.38&-3.36&-4.06&S &2,15,25\\
NGC 6240&0.0245&2&39.77&39.85&9.15 &-1.80&-2.03&-2.14&-2.64&S &7,7,42\\
NGC 6300&0.0037&2&36.82& &6.92 &-1.31&-3.24&-2.84&-2.83&A&6,25\\
NGC 6890&0.0081&2&37.25&37.29&7.07 &-1.37&-2.90&-2.72&-2.95&A&2,19,25\\
NGC 7130&0.0162&2&38.93& &7.52&-0.58&-2.45&-2.39&-2.67&S &2,31\\
NGC 7213&0.0058&1.5&39.06& &7.99&-2.86&-0.51&-0.93&-1.22&S &1,36\\
NGC 7314&0.0048&1.9&36.84& &6.03 &0.02&-3.65&-3.17&-2.84&A&2,25\\
NGC 7469&0.0163&1.5&38.89&39.17&7.09$^{e}$&0.04&-2.69&-2.43&-3.21&S &3,22,28\\
NGC 7582&0.0053&2&38.41&38.14&7.40 &-0.84&-2.60&-2.19&-2.53&A&2,21,25\\
NGC 7590&0.0053&2&$<$36& &6.90 &-1.83&-3.52&-3.32&-3.67&A&2,25\\
NGC 7603&0.0295&1.5&38.73&38.75&8.08&-1.40&-2.40&-2.40&-2.60&S &2,15,36\\
NGC 7674&0.0289&2&39.79&39.70&7.56 &-0.08&-2.14&-2.14&-1.73&A&3,23,36\\
TOL1238-364&0.0109&2&38.32&$\geq$38.45&6.83&-0.79&-2.17&-2.62&-2.68&S &2,19,31\\
UGC 7064&0.0250&1.9&$<$37.36&38.16& & &-4.01&-3.77&-3.62&A&2,15\\
UGC 12138&0.0250&1.8&38.34& &7.16&-0.43&-2.84&-2.70&-2.27&A&5,31\\
UM 146&0.0174&1.9&38.54& &6.19&-0.52&-1.58&-1.89&-1.76&S &3,31\\

\hline
\end{tabular}
\medskip
$^a$ 8.4~GHz High-resolution data was unavailable, thus we extrapolated measurements from other wavelengths assuming, $f_\nu \propto \nu^{-0.7}$ \\
$^b$ Values for the stellar velocity dispersion from the FWHM of [O~III] $\lambda~5007$, $\sigma_{[OIII]}$ \\
$^c$ Maser Kinematics\\
$^d$ Virial Theorem\\
$^e$ Reverbertation Mapping \\
$^f$ From broad ${\rm H_\alpha}$ emission\\
$^g$ $M_{BH}-\sigma_{bulge}$ relationship\\ 
Note: The luminosities were calculated from the fluxes using  $H_o=71{\rm kms^{-1}Mpc^{-1}}$ and  a deceleration 
parameter $q_o=0$ with  redshift values taken from NED. ${\rm R_{[O IV]}}$, $ {\rm R_{[Ne III]}}$ and  $ {\rm R_{[Ne II]}}$ are the 8.4GHz-to-mid-infrared parameters. The last two columns show the main contribution (MC) to the [Ne~II] emission line (A=AGN and S=Star Formation) and the references for the observed  high-resolution 8.4~GHz and 1.4~GHz emission fluxes and black holes masses ${\rm M_{BH}}$, respectively. Most of the values for the BH masses were calculated by using the relationship between the black hole mass and the stellar velocity dispersion \citep{2002ApJ...574..740T}, unless noted.\\
References: (1)  \citet{2001ApJ...555..663S}, (2)  \citet{2000MNRAS.314..573T}, (3)  \citet{1995MNRAS.276.1262K}, (4)  \citet{1984ApJ...278..544U}, (5)  Schmitt et al. 2009, in prep, (6)  \citet{1999A&AS..137..457M}, (7) \citet{1994ApJ...436...89C}, (8)  \citet{1998MNRAS.293..189D}, (9)  \citet{1994ApJ...432..496R}, (10)  \citet{1998MNRAS.300.1111H}, (11)  \citet{1998MNRAS.301.1019R}, (12) \citet{1986ApJ...310..136U}, (13)  \citet{1999ApJS..120..209N}, (14)  \citet{1987AJ.....93..255C}, (15)  \citet{1995ApJ...450..559B}, Faint Images of the Radio Sky at Twenty Centimeters (FIRST) survey, (16)  \citet{1995AJ....109...81U}, (17)  \citet{1987MNRAS.228..521U}, (18)  \citet{1995A&A...295..585S}, (19) \citet{1989ApJ...343..659U}, (20)  \citet{1984ApJ...285..439U}, (21)  \citet{1985MNRAS.216..193M}, (22)  \citet{2006AJ....131..701L}, (23)  \citet{1985ApJ...296...60M}, (24)  \citet{2003ApJ...590..162G}, (25)  \citet{2004MNRAS.355..273C},  (26)  \citet{1992ApJS...79...49W},  (27)  \citet{2007ApJ...661..660K},  (28)  \citet{2004ApJ...613..682P},  (29)  \citet{1991MNRAS.253...19L},  (30)  \citet{1996MNRAS.281.1206Y},  (31)  \citet{2005MNRAS.359..765G},  (32)  \citet{2007ApJ...661L.143W}, (33)  \citet{2007ApJS..169....1O}, (34)  \citet{1988ApJS...67..249D}, (35)  \citet{2007ApJ...668...94H}, (36)  \citet{1995ApJS...99...67N}, (37)  \citet{2007ApJ...667..131G}, (38)  \citet{2005MNRAS.358.1405O}, (39)  \citet{1990AJ.....99.1722K}, (40)  \citet{2009MNRAS.393L...1R},  (41)  \citet{2009ApJS..183....1H},  (42)  \citet{1994ApJ...437L..23D}   

\end{minipage}
\end{table*}

\begin{figure}
\includegraphics[width=84mm]{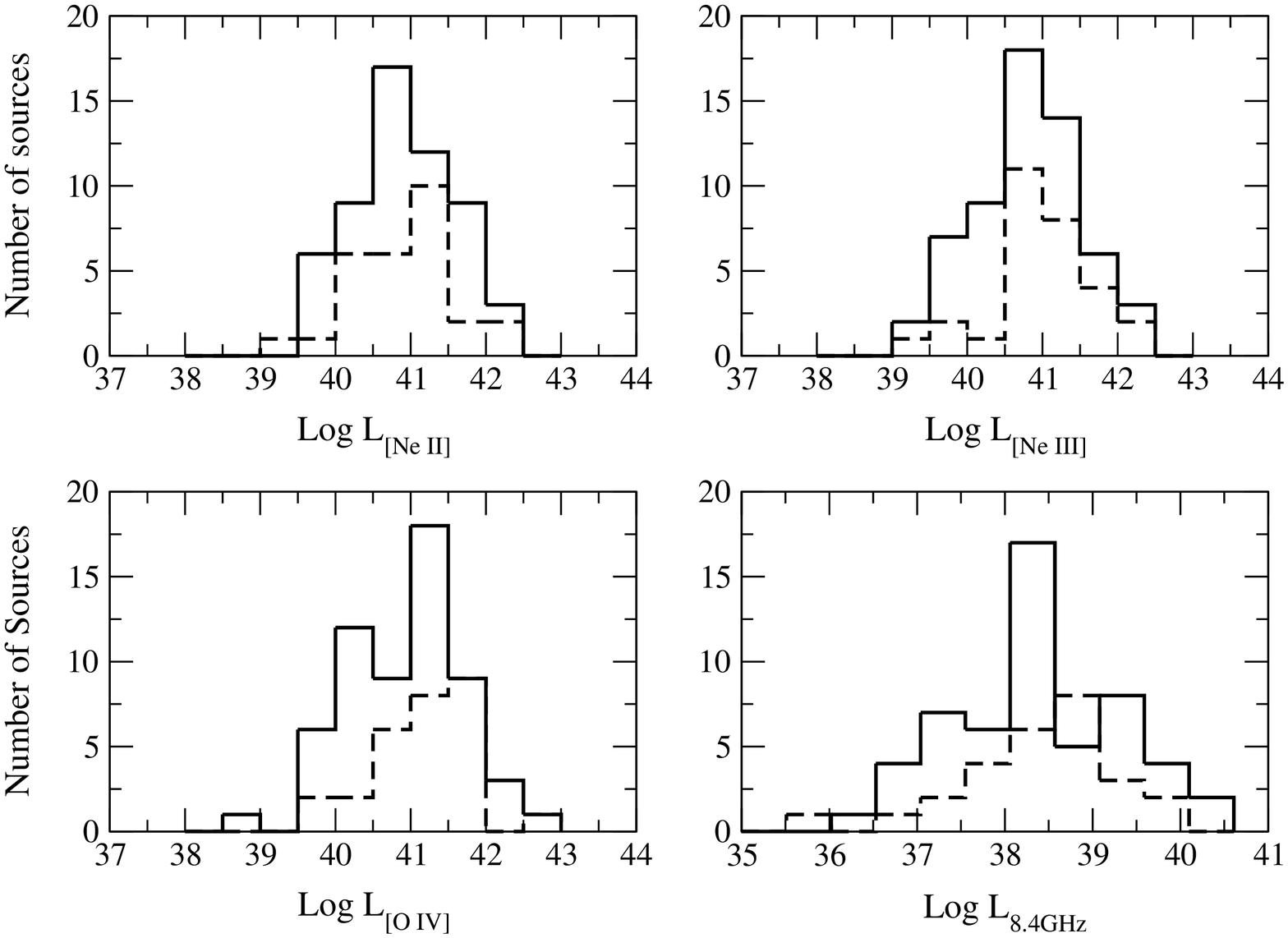}
    \caption{Comparison of the [Ne~II], [Ne~III], [O~IV] and 8.4~GHz luminosities (in $ergs~s^{-1}$) in Seyfert 1 ({\it dashed line}) and Seyfert 2 ({\it solid line}) galaxies. This sample includes 29 Seyfert 1 galaxies and 59 Seyfert 2 galaxies for the mid-infrared luminosity  distributions. Upper limits are not included in the 8.4~GHz luminosity distribution. The K-S test for these luminosities  returns a $\sim 96.9\%$, $\sim 47.9\%$,$\sim 27.8\%$ and $\sim 82.0\%$ probability of the null hypothesis for the  [Ne~II], [Ne~III], [O~IV] and 8.4~GHz luminosities, respectively \label{hist}}
\end{figure}

In order to confirm that our sample is not biased in terms of luminosities we compared  the different luminosities  for both the mid-infrared emission lines and the radio continuum luminosities for the two groups of galaxies. The histograms of [Ne~II], [Ne~III] and [O~IV]  luminosities and 8.4~GHz continuum luminosities  are presented in  Figure~\ref{hist} with the results from the  Kolmogorov-Smirnov (K-S) test presented in Table~\ref{table2}. This table also includes information about the numbers of Seyfert 1 and Seyfert 2 galaxies, median values and standard deviations of the mean for the measured quantities. For the 8.4~GHz continuum the K-S test  returns a $\sim 82.0\%$ probability of the null hypothesis  (i.e., that there is no difference between Seyfert 1 and Seyfert 2 galaxies), or in other words,  two samples drawn from the same parent population would differ this much $\sim 82.0\%$ of the time\footnote{A probability  value of less than $5\%$ represents a 2$\sigma$ level of significance that two samples drawn from the same population are different. A strong level of significance, at 3$\sigma$ level, is obtained for values smaller than $1\%$ \citep[e.g.,][]{1992nrfa.book.....P,2003drea.book.....B}}. For the [Ne~II], [Ne~III] and [O~IV]  luminosity distributions the K-S test  returns a $\sim 96.9\%$, $\sim 47.9\%$ and $\sim 27.8\%$ probability of the null hypothesis, respectively. 

It can be seen that Seyfert 1 and Seyfert 2 galaxies have similar distribution of values. As discussed in  \citet{2008ApJ...689...95M}, there is an apparent  lack of low luminosity Seyfert 1 galaxies in the [O~IV] and [Ne~III] distribution, but we find this difference not to be statistically significant, as shown by the K-S test results. Moreover, the sample presented in this work spans the same range on [O~IV] luminosities as the revised Shapley-Ames sample \citep{2009ApJ...698..623D}. The good agreement between Seyfert 1 and Seyfert 2  galaxies in the radio continuum and mid-infrared emission line  luminosity distributions demonstrates the isotropic properties of these quantities.

\begin{table*}
 \centering
 \begin{minipage}{140mm}
\caption{Statistical Comparison  Between Seyfert 1 and Seyfert 2 Galaxies\label{table2} 
}
 \begin{tabular}{@{}lccccccc@{}}
  \hline
& \multicolumn{3}{c}{Seyfert 1}& \multicolumn{3}{c}{Seyfert 2}\\
 \cline{2-4} \cline{5-7}\\
 &  Measurements && Standard & Measurements &&Standard &${\rm P_{K-S}}$\\
 & Available & Median&Deviation &Available&Median&Deviation& (\%)\\
 \hline

{\rm  Log 8.4GHz/[O~IV]} &27&  -2.8  & 0.1  &55&  -2.6   &  0.1 &  8.2   \\
{\rm Log 8.4GHz/[Ne~III]} &27 & -2.6 &  0.1 &55 & -2.5 &  0.1 &  57.3 \\
{\rm Log 8.4GHz/[Ne~II]} & 27 & -2.5 & 0.1  & 55&  -2.6  & 0.1 &  81.6\\

${\rm L_{1.4GHz}}$ & 19  & 38.3 & 0.2 &36 &38.2 &0.1 & 86.8\\
${\rm L_{8.4GHz}}$ &27& 38.4& 0.2& 55  & 38.4& 0.1 & 82.0\\
${\rm L_{[O IV]}}$ & 29& 41.3  &0.1  &59  & 41.0& 0.1 &  27.8\\
${\rm L_{[Ne III]}}$ &29 &41.0& 0.1& 59 & 40.9  & 0.1 & 47.9 \\
${\rm L_{[Ne II]}}$ &29 & 40.9 &0.1 & 59& 41.0 & 0.1 & 96.9  \\

\hline
\end{tabular}
\medskip
Note: The last column, ${\rm P_{K-S}}$,  represents the Kolmogorov-Smirnov (K-S) test null probability. Upper limits for the  8.4~GHz fluxes are not included.
\end{minipage}
\end{table*}

\section{Mid-Infrared and the 8.4~GHz (3.6~cm) Emission}
In order to understand the different contributions to the radio core emission and its interaction with the narrow-line region in Seyfert galaxies we investigated  the  mid-infrared emission line and 8.4~GHz correlations. One should note that, due to redshift effects, luminosity-luminosity  plots will almost always show some correlation. Furthermore, caution must be taken when applying statistical analysis to data sets that contain non-detections (upper limits), or  ``censored" data
 points. In order to deal with these problems we used Astronomy SURVival analysis {\bf ASURV} Rev 1.2 \citep{1990BAAS...22..917I,1992BAAS...24..839L}, which implement the methods presented in  \citet{1986ApJ...306..490I}. We also used a test for partial correlation with censored data \citep{1996MNRAS.278..919A} in order to exclude the redshift effect in the correlations. A detailed comparison between the correlation coefficient for different  radio-to-mid-infrared correlations  is presented in Table~\ref{table3}.

\begin{table*}
 \centering
 \begin{minipage}{140mm}
\caption{Statistical Comparison of the   Relationships Between Radio and Mid-Infrared Emission Lines Luminosities for the  Sample\label{table3}}
 \begin{tabular}{@{}lccccccc@{}}
  \hline
$\log - \log$   & $\rho_s$& $P_\rho$&  $\tau$&  $P_\tau$& $\tau_p$& $\sigma$& $P_{\tau,p}$\\
 \hline

{\bf All The Sample}\\
8.4GHz-[O~IV]&0.66&$<1\times 10^{-6}$&0.50&$<1\times 10^{-6}$&0.35&0.08&$<1\times 10^{-6}$\\
8.4GHz-[Ne~III]&0.75&$<1\times 10^{-6}$&0.57&$<1\times 10^{-6}$&0.44&0.07&$<1\times 10^{-6}$\\
8.4GHz-[Ne~II]&0.74&$<1\times 10^{-6}$&0.54&$<1\times 10^{-6}$&0.40&0.06&$<1\times 10^{-6}$\\

{\bf AGN Dominated Sources}\\
8.4GHz-[O~IV]&0.76&$<1\times 10^{-6}$&0.58&$<1\times 10^{-6}$&0.46&0.09&$<1\times 10^{-6}$\\
8.4GHz-[Ne~III]&0.80&$<1\times 10^{-6}$&0.62&$<1\times 10^{-6}$&0.51&0.09&$<1\times 10^{-6}$\\
8.4GHz-[Ne~II]&0.82&$<1\times 10^{-6}$&0.64&$<1\times 10^{-6}$&0.53&0.09&$<1\times 10^{-6}$\\

{\bf Star Formation Dominated Sources}\\
8.4GHz-[O~IV]&0.59&$1.30\times 10^{-3}$&0.42&$5.00\times 10^{-4}$&0.22&0.12&$6.23\times 10^{-2}$\\
8.4GHz-[Ne~III]&0.68&$1.92\times 10^{-4}$&0.51&$2.80\times 10^{-5}$&0.32&0.12&$8.80\times 10^{-3}$\\
8.4GHz-[Ne~II]&0.64&$5.00\times 10^{-4}$&0.47&$1.00\times 10^{-4}$&0.30&0.11&$6.60\times 10^{-3}$\\

\hline
\end{tabular}
\medskip
Note:  $\rho_s$ is the Spearman rank order correlation coefficient with its associated null probability, $P_\rho$.
$\tau$ represents the generalized Kendall's correlation coefficient for censored data and $\tau_p$ is the  Kendall's  coefficient for partial correlation with censored data. $P_\tau$ and $P_{\tau,p}$ are the  null probabilities for the generalized and partial Kendall's correlation test, respectively. We also show the calculated variance, $\sigma$, for Kendall $\tau_p$. We have used a partial correlation test to exclude the effect of redshift (distance) in the luminosity-luminosity correlations. 
\end{minipage}
\end{table*}

\begin{figure}
\includegraphics[width=84mm]{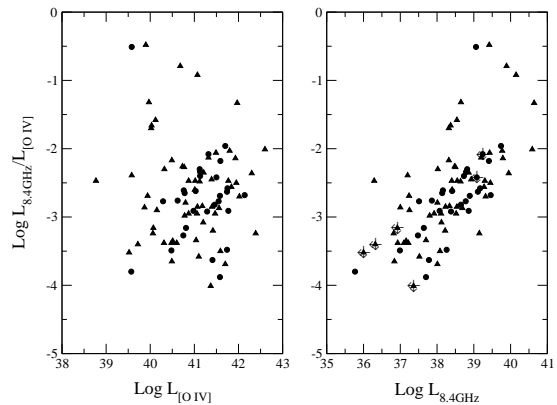}
    \caption{Correlation between the  ratio {\rm $L_{8.4GHz}/L_{[O IV]}$} and the [O IV] and 8.4~GHz luminosities in Seyfert 1 ({\it circles}) and Seyfert 2 ({\it triangles}) galaxies. \label{fig5_5}}
\end{figure}

 We first considered the ratio between the radio continuum at 8.4~GHz ($L_{8.4GHz}$) and the [O~IV] emission line. The K-S test for this ratio returns a $\sim 8.2\%$ probability of the null hypothesis (see Table~\ref{table2}). Therefore, the radio emission relative to the power of the AGN, as measured by the [O~IV] emission,  is statistically similar in both Seyfert 1 and Seyfert 2 galaxies, suggesting the isotropic property of the radio emission \citep[e.g.,][]{2001MNRAS.325..737T,2009ApJ...698..623D}. This result is expected because the radio emission is expected to be unaffected by the  torus. When using the generalized Spearman correlation test to compare this ratio with the [O~IV] and 8.4~GHz luminosities we find no correlation with the [O~IV] luminosities  ($\rho_s=0.05;~P_\rho=0.64$) and a good correlation with the 8.GHz luminosity ($\rho_s=0.72;~P_\rho<1\times 10^{-6}$), see Figure~\ref{fig5_5}. These results suggest that the radio loudness of the source cannot be associated solely with  the strength of the AGN. In the following section we will elaborate on the implications of this result in view of  the link between radio loudness and the mass accretion rates for the AGN. From the  8.4GHz/[Ne~III] and 8.4GHz/[Ne~II] ratios we found that Seyfert 1 and Seyfert 2 galaxies also have statistically similar distributions. The K-S test returns a  57.3$\%$ and 81.6$\%$ probability of the null hypothesis for the  8.4GHz/[Ne~III] and 8.4GHz/[Ne~II] ratios, respectively (see Table~\ref{table2}). The 8.4GHz-to-mid-infrared ratios are presented in Table~\ref{table1}.

Because Seyfert~1 and Seyfert~2 galaxies have statistically similar radio emission, relative to the strength of the AGN, we   separated our sample in two new groups: 1) AGN  and 2) star formation dominated sources, by following our previous  study on the AGN and star formation contribution to the [Ne~II] emission line  \citep{2008ApJ...689...95M}. In this new scheme, sources with more than 50$\%$ contribution from the AGN to the [Ne~II] emission line are considered to be AGN dominated and sources with less than 50$\%$ contribution from the AGN to the [Ne~II] emission line, or alternatively, more than 50$\%$ contribution from stellar activity to the [Ne~II] emission  are star formation dominated. One must note that in active galaxies with strong star formation, the [O~IV] emission could be contaminated by the stellar activity, where the most likely contributor to the [O~IV] emission is from HII regions ionized by  Wolf-Rayet stars \citep[see][for discussion]{2009ApJ...704.1159H}. However, if we correct the [O~IV] emission for stellar contamination in star formation dominated sources, the resulting [O~IV]/[Ne~II] ratio will have even smaller values, thus reinforcing this quantitative classification. Moreover, sources that are well known to harbor massive star formation regions, such as NGC~3079 and NGC~6240 may have an additional complication because of contamination from the adjacent [Fe~II] $\lambda$25.99$\mu {\rm m}$ line from the [O~IV] in low-resolution  IRS spectra. Following the discussion  presented in  \citet{2008ApJ...682...94M} for NGC~3079, we estimated the  starburst contribution to be  $\sim 1.5$ times the uncontaminated [O~IV] emission; similar results are obtained  when comparing our fluxes with high-resolution spectroscopy for NGC~6240 \citep{2006ApJ...640..204A}. Nevertheless, these  extreme cases  have been classified correctly as star formation dominated sources. Moreover, a factor of $\sim 2$ in the [O~IV] flux  represents a dispersion of 0.3~dex, thus within the observed dispersion for the correlations presented in this study.

\begin{figure}
\includegraphics[width=84mm]{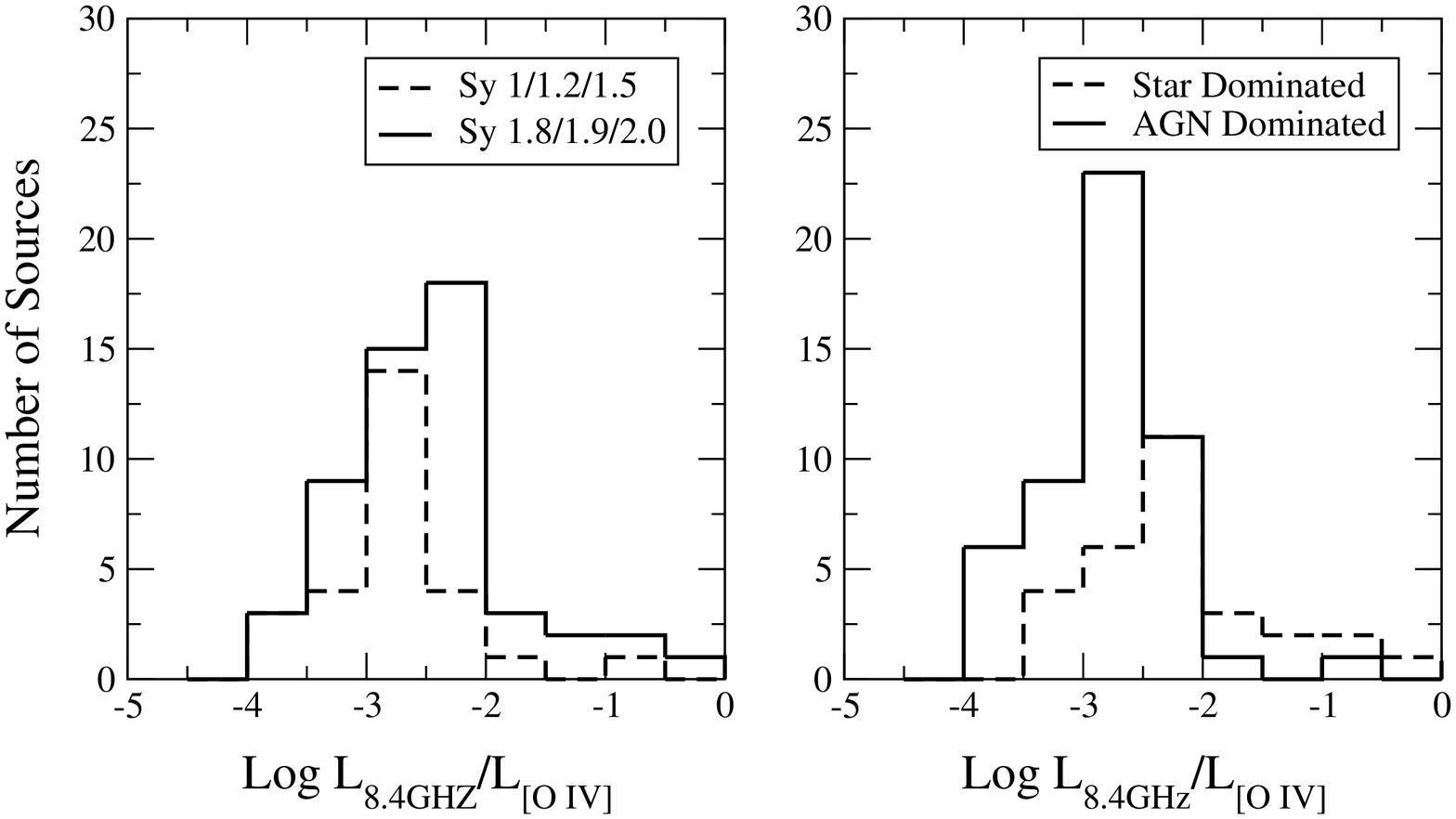}
    \caption{Comparison of the {\rm $L_{8.4GHz}/L_{[O IV]}$} ratio in Seyfert 1 and Seyfert 2 galaxies  and AGN and star formation dominated sources. This sample includes 53 AGN dominated  Seyfert galaxies and 29 star formation dominated sources for the  distributions. Upper limits for the  8.4~GHz fluxes are not included. The K-S test for these  ratios return  $\sim 8.2\%$ and $\sim 0.1\%$ probability of the null hypothesis between Seyfert 1 and Seyfert 2 galaxies and AGN and star formation dominated galaxies, respectively \label{hist_1}}
\end{figure}

 From this new classification we found that, star formation and AGN dominated sources are statistically different in their radio emission relative to the AGN. The K-S test for the 8.4GHz/[O~IV] ratio returns a $\sim 0.1\%$ probability of the null hypothesis, i.e., indicating that  star formation and AGN dominated galaxies are from different populations. Figure~\ref{hist_1} shows  the distribution of values for the {\rm $L_{8.4GHz}/L_{[O IV]}$} ratio for Seyfert 1 and Seyfert 2 galaxies, as well as for AGN and star formation dominated sources, where  the differences in distributions can be clearly seen.  On the other hand, the K-S test returns a  11.9$\%$ and 10.2$\%$ probability of the null hypothesis for the  8.4GHz/[Ne~III] and 8.4GHz/[Ne~II] ratios, respectively (see Table~4). Table~4 also includes information about the numbers of AGN and star formation dominated sources, median values and standard deviations of the mean for the measured quantities. One must note that a high stellar contribution to the [Ne~II] emission, based on the [O~IV]/[Ne~II] ratio and typical of  star formation dominated sources, does not always translate into higher SFRs, but  could also indicate a relatively weak AGN \citep[see][]{2008ApJ...689...95M}. 

\begin{table*}
 \centering
 \begin{minipage}{140mm}
\caption{Statistical Comparison Between AGN and Star Formation  Dominated Sources\label{table4}}
 \begin{tabular}{@{}lccccccc@{}}
  \hline
& \multicolumn{3}{c}{AGN Dominated }& \multicolumn{3}{c}{STAR Dominated}\\
 \cline{2-4} \cline{5-7}\\
 &  Measurements && Standard & Measurements &&Standard &${\rm P_{K-S}}$\\
 & Available & Median&Deviation &Available&Median&Deviation& (\%)\\
 \hline

{\rm  Log 8.4GHz/[O IV]} &53 &  -2.8  &  0.1  & 29&  -2.4 & 0.1   &  0.3   \\
{\rm Log 8.4GHz/[Ne III]} &53  &-2.7  &   0.1 & 29 & -2.4 &   0.1&   11.9 \\
{\rm Log 8.4GHz/[Ne II]} &  53 &  -2.5   & 0.1   & 29 &   -2.7  &  0.1& 10.2  \\

{\rm  Log 1.4GHz/[O IV]} &27 &  -3.0   &  0.1  & 16&  -2.5& 0.1   &  0.2   \\
{\rm Log 1.4GHz/[Ne III]} &27  &-2.8   &   0.1 & 16 & -2.4 &   0.1&   18.9 \\
{\rm Log 1.4GHz/[Ne II]} &  27 &  -2.8  & 0.1   & 16 &   -2.9  &  0.1& 12.3  \\

\hline
\end{tabular}
\medskip
Note: The last column, ${\rm P_{K-S}}$,  represents the Kolmogorov-Smirnov (K-S) test null probability.  Upper limits for the  8.4~GHz and 1.4~GHz fluxes are not included.
\end{minipage}
\end{table*}

\section{The Radio-to-Mid-Infrared radio loudness diagnostic}

\begin{figure*}
\vbox to 174mm{\vfil  \includegraphics[width=174mm]{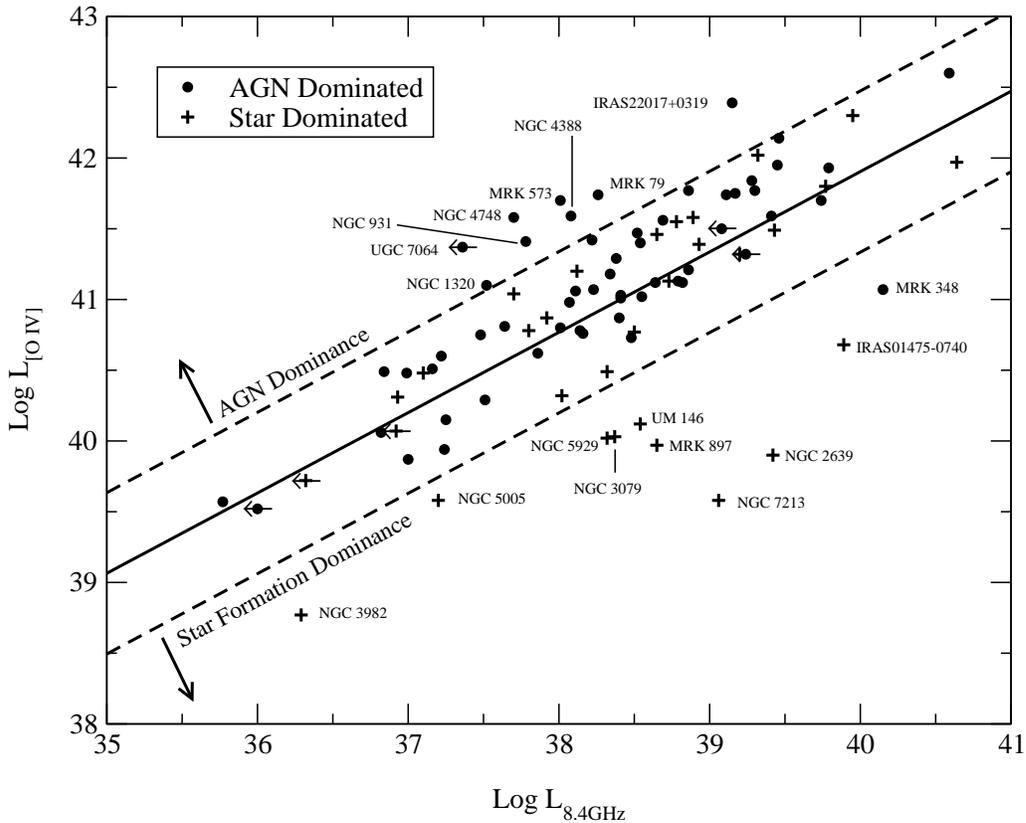}
    \caption{Correlation between the [O~IV] and the 8.4~GHz emission. The circles represent the AGN dominated sources and the crosses  represent the star formation dominated sources. The solid line represents the linear regression for the whole sample and the dashed lines represent a $1-\sigma$ dispersion. \label{fig1_3}}
\vfil}
\end{figure*}

\begin{figure}
\includegraphics[width=84mm]{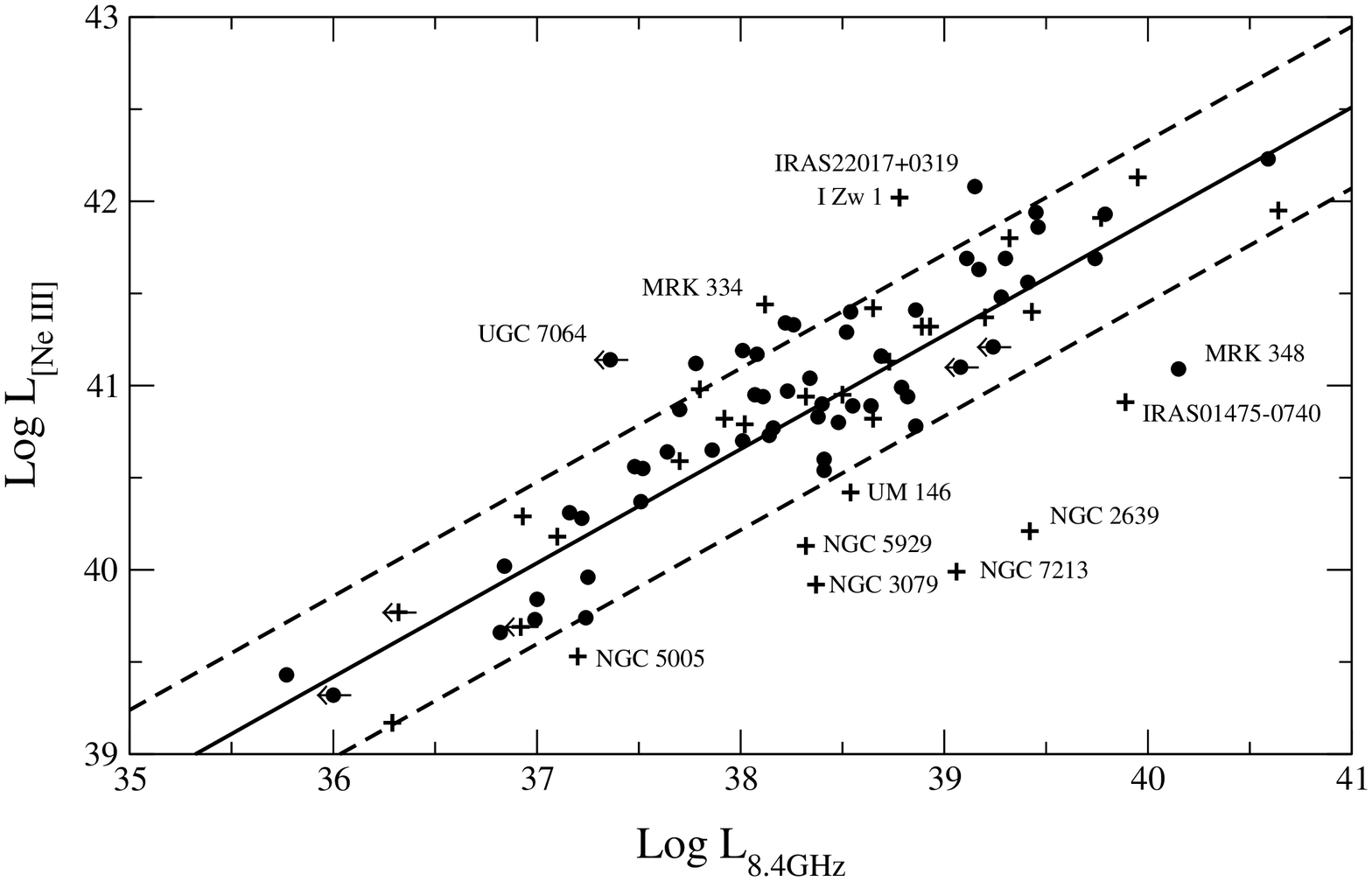}
    \caption{Correlation between the [Ne~III] and the 8.4~GHz emission. The circles represent the AGN dominated sources and the crosses  represent the star formation dominated sources. The solid line represents the linear regression for the whole sample and the dashed lines represent a $1-\sigma$ dispersion \label{fig1_4}}
\end{figure}

\begin{figure}
\includegraphics[width=84mm]{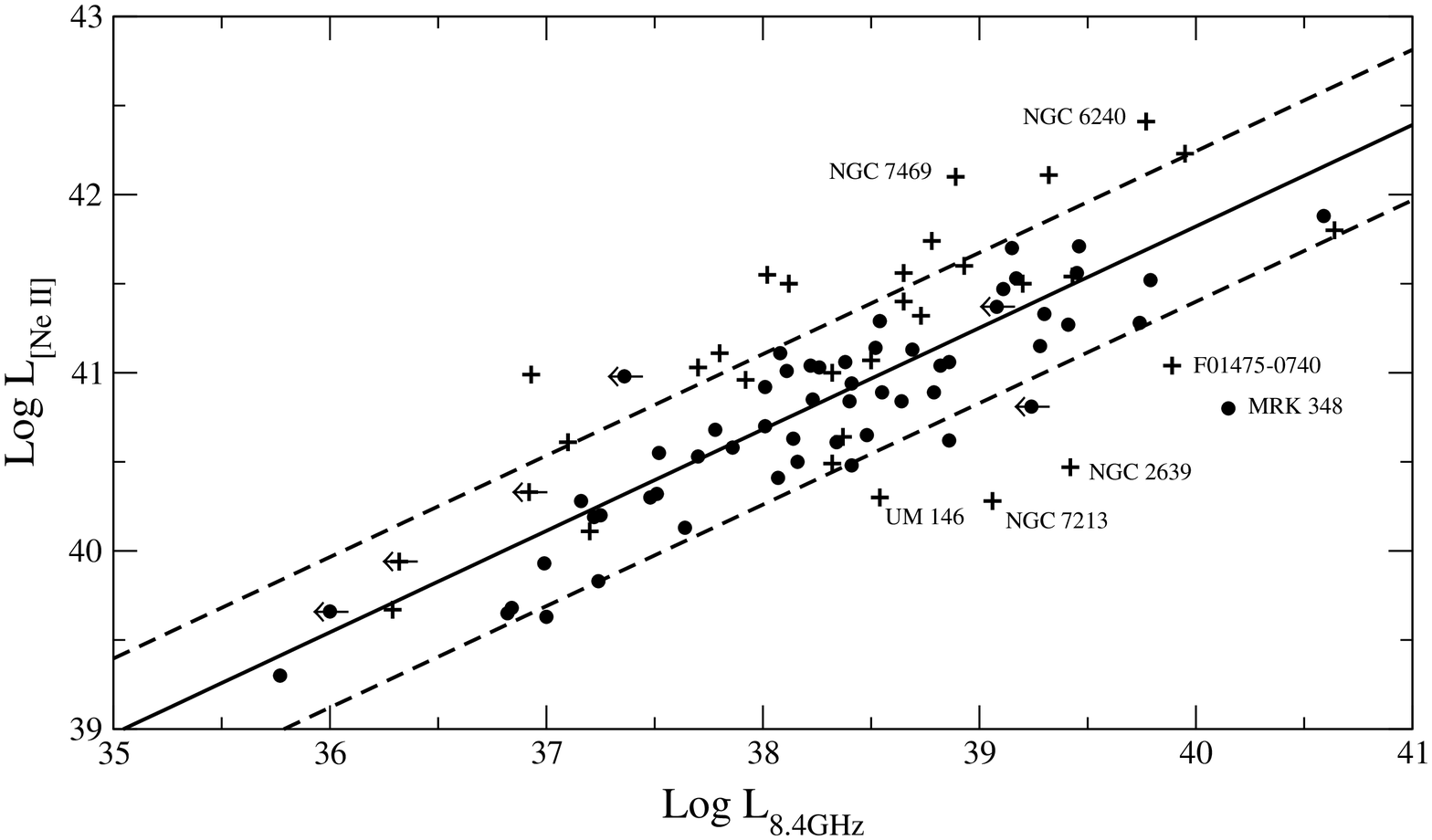}
    \caption{Correlation between the [Ne~II] and the 8.4~GHz emission.The circles represent the AGN dominated sources and the crosses  represent the star formation dominated sources.  The solid line represents the linear regression for the whole sample and the dashed lines represent a $1-\sigma$ dispersion\label{fig1_5}}
\end{figure}

 By looking at the mid-infrared and radio correlations in terms of the dominant source of  the [Ne~II] emission, we can  investigate on the relationship between the radio emission and the nuclear activity. Figures~\ref{fig1_3}-\ref{fig1_5} show the mid-infrared and 8.4~GHz luminosities with AGN and star formation dominated sources indicated.  In particular,  in the [O~IV]-8.4~GHz correlation, Figure~\ref{fig1_3}, the upper region of the plot ($>+\sigma$) contains only AGN dominated sources while star formation dominated sources tend to occupied the lower region ($<-\sigma$). Comparing Figures~\ref{fig1_3}-\ref{fig1_5} one must note how the clear separation between the two groups is lost when decreasing the ionization potential of the emission lines or, alternatively, when the emission line has a stronger starburst component, in agreement with the K-S test results. For the star formation dominated region, the [O~IV] underpredicts the observed 8.4~GHz emission. This suggest two possible scenarios: 1) star formation could enhance the 8.4~GHz emission and/or 2) the AGN in star formation dominated sources  is  more radio loud than estimated via standard radio-to-optical estimates.

 In order to determine  if  star formation activity  may contaminate the nuclear 8.4~GHz emission we proceed to estimate the 8.4~GHz emission associated with the star formation rates (SFR). For the SFR we adopted the values derived from  the stellar component of the [Ne~II] emission \citep{2008ApJ...689...95M}. Then  we used  the calibration found by  \citet{2006ApJ...643..173S} between the SFR and the 8.4~GHz emission. This calibration includes both the thermal and non-thermal SFR radio contribution (see Table~\ref{table5}). From this we found that the SFR overpredicts the observed 8.4~GHz luminosities in NGC~3982, NGC~5005 and MRK~897. This result is expected because the VLA in A configuration at 8.4~GHz sample a more compact region than that from the short-low and short-high resolution {\it Spitzer} slits, from which we extracted the [Ne~II] emission. Therefore, the 8.4~GHz emission derived from the [Ne~II] SFRs, represents an    upper limit for the star formation contribution, if any, to the radio emission at 8.4~GHz. On the other hand, the SFR underpredicts the 8.4~GHz emission for 6 out of 9 of the sources in the star formation dominated region. For these six sources, stellar activity cannot be solely responsible for the observed radio excess because it can only account for a contribution as little as $\sim 2\%$ and no greater than $\sim 40 \%$ of the observed flux, even when including the radio contribution from the extended star formation associated with the [Ne~II] emission. One must note that the estimated SFRs  represent a lower-limit as we neglect the stellar [Ne~III] contribution \citep[see][for discussion]{2008ApJ...689...95M}. However, all these sources have  weak [Ne~III], compared to the [Ne~II] emission, hence the SFR will not be significantly affected nor the estimates for the radio emission.

\begin{table}
\caption{Comparison between the observed and predicted 8.4~GHz luminosities for the sources in the star formation dominated  region\label{table5} }
 \begin{tabular}{@{}lll@{}}
  \hline
Name & Obs.& Pred.(SFR)\\
            &\multicolumn{2}{c}{Luminosity (${\rm W~Hz^{-1}}$)}\\
 \hline
IRAS01475-0740&22.96&21.38\\
MRK 897&21.73&22.02\\
NGC 2639&22.50&20.83\\
NGC 3079&21.44&21.02\\
NGC 3982&19.37&20.06\\
NGC 5005&20.27&20.44\\
NGC 5929&21.40&20.84\\
NGC 7213&22.14&20.66\\
UM 146&21.61&20.49\\

\hline
\end{tabular}
\medskip

Note: In order to estimate the predicted 8.4~GHz luminosities  we used the SFR values derived from  the stellar component of the [Ne~II] emission \citep{2008ApJ...689...95M}. We adopted the calibration found by  \citet{2006ApJ...643..173S} between the SFR and the 8.4~GHz emission. 
\end{table}

 \citet{2002ApJ...564..120H} found an intriguing anticorrelation between the radio loudness of the galaxies, given by the radio-to-optical parameter, and the Eddington ratios, ${\rm L_{bol}/L_{Edd}}$, where $L_{bol}$ is the bolometric luminosity and $L_{Edd}$ is the Eddington luminosity\footnote{The Eddington luminosity is calculated as $1.26 \times 10^{38} M/M_\odot~{\rm erg~s^{-1}}$}. In general he found that galaxies with Eddington ratios $<10^{-5}$ are in the radio loud territory, while galaxies with Eddington ratios of $>1$ are in the radio quiet regime. In order to investigate this dichotomy    we  first need to estimate ${\rm L_{bol}}$. \citet{2004ApJ...613..109H}  first suggested that  the luminosity of [O~III]~5007\AA \ \ can be use as a proxy for the bolometric luminosity. In this regard, and by extending this analysis into the mid-infrared, \citet{2008ApJ...674L...9D} found that ${\rm L_{[O~IV]}}$ is related to  the 5100\AA \ \  continuum luminosity for a sample of reverberation-mapped AGN. From this and by using an estimate for the  correction between the bolometric luminosity and the optical luminosity, ${\rm L_{bol} \approx 9\lambda L_\lambda(5100)}$ \citep{2000ApJ...533..631K}, they determined  a scaling factor of 4500 between ${\rm L_{bol}}$ and  ${\rm L_{[O~IV]}}$ with a variance of 0.5~dex. Because [O~IV] is less sensitive to reddening corrections and  is less contaminated by star-forming regions, by  using ${\rm L_{[O~IV]}}$ as a bolometric luminosity indicator we can avoid the limitations of the ${\rm L_{[O~III]}}$ method. However, caution must be taken as for any of these methods the assumption is that the luminosity from narrow emission lines represents the same fraction of ${\rm L_{bol}}$ for all AGN. Moreover, the bolometric correction could be a function of the source luminosity and/or redshift \citep{2006A&A...453..525N}.

 In order to calculate the Eddington luminosity we searched the literature for  black hole (BH) masses, and found  81 BH masses for our radio sample. For the sake of uniformity, most of the values for the BH masses were calculated by using the relationship between the black hole mass and the stellar velocity dispersion found  by  \citet{2002ApJ...574..740T}, except when noted in Table~\ref{table1}. Previous studies have suggested  an uncertainty of 0.3~dex for the BH masses \citep[e.g.,][]{2002ApJ...564..120H}, however a more conservative uncertainty of 0.5~dex may be more appropriate for relatively low mass BH ($M_{BH}<10^7 M_\odot$). One must note that the range for the normalized accretion rate, ${\rm L_{bol}/L_{Edd}}$, in our sample is in agreement with the range of values found for the  radio quiet sources presented in  \citet{2002ApJ...564..120H}. The Eddington ratios, ${\rm L_{bol}/L_{Edd}}$,  are presented in Table~\ref{table1}.

\begin{figure}
\includegraphics[width=84mm]{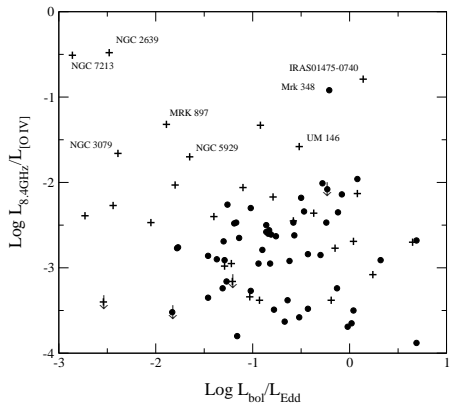}
    \caption{Comparison between ${\rm L_{bol}/L_{Edd}}$ and ${\rm L_{8.4GHz}/L_{[O IV]}}$ as a proxy for the AGN radio loudness. The symbols are the same as in Figure~\ref{fig1_3}\label{edd_ratio}}
\end{figure}

\begin{figure}
\includegraphics[width=84mm]{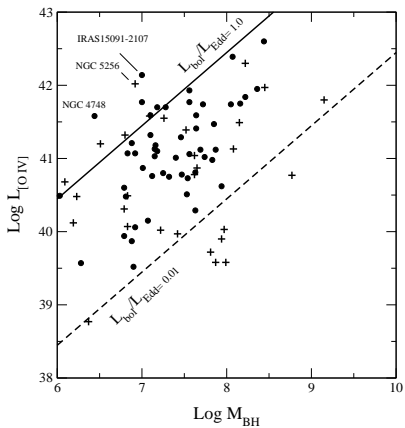}
    \caption{Dependence of [O~IV] luminosity on black hole mass. The solid and dashed lines represent different values for Eddington accretion rates.  The symbols are the same as in Figure~\ref{fig1_3}\label{comp_o4_BHM}}
\end{figure}

In Figure~\ref{edd_ratio} we show that a number of  star formation dominated sources have smaller Eddington ratios in agreement with the radio excess observed in the star formation  dominated  region in the [O~IV]-8.4~GHz correlation. There is certainly a trend of  the Eddington ratios decreasing with increase radio loudness for star formation dominated sources, but  this correlation is  not statistically significant with this trend been dominated by the   sources  in the star formation dominated region. Also, there is an inherent scatter because of the uncertainties in the  BH mass estimate. Nevertheless, these results show  that star formation dominated sources with the bigger radio contribution, i.e., sources in the star formation dominated region, have smaller mass accretion rates. In this regard we also examined the dependence of the [O~IV] luminosity on black hole mass (see Figure~\ref{comp_o4_BHM}). From this comparison it can be seen that star formation dominated sources tend to have smaller accretion rates. Although the scatter is very large, there is a trend of $L_{[O IV]}$ increasing with $M_{BH}$.

Using the radio-to-mid-infrared  ratio,  8.4GHz/[O~IV],   we found that star formation dominated sources are, on average, $\sim 2.8$ times more radio loud than AGN dominated sources (see Table~\ref{table4}). This ratio has the advantage of classifying galaxy radio loudness without star formation contamination and/or dust extinction in the determination of the optical luminosity. On the other hand, the galaxy radio loudness determined from the 8.4GHz/[Ne~III] and  8.4GHz/[Ne~II] ratios could be a closer match   to the historically  radio loudness parameter, $R$, because these  fine-structure lines of neon could be  affected by star formation contamination, especially in weak or low-luminous AGN. Moreover, in some AGN, the observed   [Ne~V]14.32~$\micron$/[Ne~V]24.32~$\micron$  ratio seems to fall below the low-density limit, implying a wavelength dependent extinction in the mid-infrared, with stronger extinction at shorter wavelengths \citep[e.g.,][]{2007ApJ...664...71D,2009MNRAS.398.1165G}, suggesting that [Ne~III] and  [Ne~II] could be more affected by dust extinction than [O~IV],  attributing  the   statistical differences between radio loudness in AGN and star formation dominated sources observed with  these ratios.

Overall, we found a good correlation between all the mid-infrared lines and the 8.4~GHz luminosities in AGN dominated sources.  On the other hand, the correlations for  star dominated sources seems to be strongly affected by the distance as these correlations are not statistically significant when using a partial correlation test with the effect of redshift excluded (see Table~\ref{table3}). These results suggest that the nuclear 8.4~GHz radio emission cannot be associated directly with the intrinsic power of the AGN, as measured with [O~IV], in sources that are dominated by star formation, or alternatively in low-luminosity AGN,  nor can it be associated with star formation emission, because the SFRs for most of these sources  underpredict the observed 8.4GHz luminosity (see Table~\ref{table5}). This may  argue in favor of an advection dominated accretion flow (ADAF) model   for the accretion disk \citep[e.g.,][]{1994ApJ...428L..13N}. In this model, the low accretion rates for the infalling material may never form a thin disk because of inefficient cooling, with the main  coolant process being cyclo-synchrotron emission, which result in a compact radio emission source that is faint in other wavelengths, providing an alternative to low-power jets models in radio quiet sources. The accretion rates for some   sources in the star formation dominated region are  sufficiently sub-Eddington to support the ADAFs accretion models,e.g.,  low-luminous sources with a bright nuclear radio emission (see Table~\ref{table1}).  Previous studies have suggested ADAF-type accretion flow models to explain the sub-Eddington accretion nature of  two of the more extreme sources  presented in this work, NGC~2639 and NGC~7213 \citep[e.g.,][]{2003ApJ...583..145T,2005MNRAS.356..727S}. Our results support  this scenario in light of the observed radio ``excess", compared to the intrinsic power of the AGN, as these  are the most  radio loud sources  in our sample (see Table~\ref{table1} and Figure~\ref{edd_ratio}). One may consider that star formation dominated sources could have a greater amount of fueling material in their nuclear regions, contrary to the ADAFs accretion models,  but as we have mentioned before, sources with higher star formation contributions to [Ne~II] do not necessarily have high SFR.

\begin{figure}
\includegraphics[width=84mm]{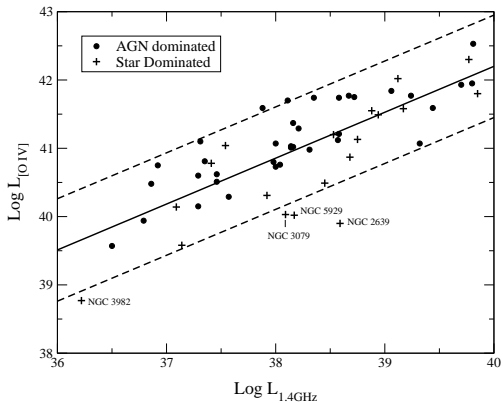}
    \caption{Correlation between the [O~IV] and the 1.4~GHz emission. The red diamonds represent the AGN dominated sources and the blue squares represent the star formation dominated sources. The solid line represents the linear regression for the whole sample and the dashed lines represent a $1-\sigma$ dispersion\label{fig2_1}}
\end{figure}

Finally, we  compared  the  radio-to-mid-infrared parameter by using VLA B-configuration radio flux densities  at 1.4~GHz and the mid-infrared emission lines, between AGN and star formation dominated sources. As we mentioned before,  to estimate the radio contribution from the AGN one needs to use high angular resolution data, which are less contaminated by star formation from  the host galaxy, except for the nuclear stellar activity. However, caution must been taken because  at 1.4~GHz the region sampled by the VLA B-configuration could be contaminated by star formation in the circumnuclear region (within the inner few kpc). However, when comparing the shape of the SED between  1.4~GHz and 8.4~GHz emission we found that  the K-S test for the spectral indices distribution, between AGN and star formation dominated sources,  returns a $\sim 51.1\%$ probability of the null hypothesis. This result suggest that stellar activity is not the leading effect that drives the radio core  correlation, as we found previously with the 8.4~GHz emission or, in other words, the high-resolution radio images at 1.4~GHz and 8.4~GHz are, on average,  dominated by the active nuclei. Moreover, \citet{2008ApJ...682...94M} suggested that because its high ionization potential the [O~IV] emission is well within the Spitzer aperture on scales of less than a few hundreds of parsecs, typical of the NLR. From  the 1.4GHz/[O~IV] ratios  we  found that star formation and AGN dominated sources are statistically different in their radio loudness. The K-S test for this ratio returns a $\sim 0.2\%$ probability of the null hypothesis with star formation dominated sources more radio loud by a factor $\sim 3$ than AGN dominated sources, in close agreement with our previous finding using the radio core emission at 8.4~GHz. On the other hand, the K-S test returns a  18.9$\%$ and 12.3$\%$ probability of the null hypothesis for the  1.4GHz/[Ne~III] and 1.4GHz/[Ne~II] ratios, respectively (see Table~4). These results corroborate our assumption that the [O~IV] emission is a true indicator of the AGN power, as the [Ne~II] and [Ne~III] emission could have a strong starburst component in weak AGN. Figure~\ref{fig2_1} shows the 1.4GHz-[O~IV] correlation, where one  can see the  separation between AGN and star formation dominated sources, in agreement with the K-S test results and  despite the smaller sample because of the  limited number of high-resolution observations.

\section{Conclusions}
We have investigated the relationship between the radio emission and the ionization state of the emission-line gas in Seyfert galaxies with the aim to understand the connection between the radio loudness and the  ionizing luminosities of the AGN. We used the [O~IV] emission line  to estimate the intrinsic power of the AGN and high-resolution 8.4~GHz emission to characterize the radio core power. From this we defined a radio-to-mid-infrared parameter, 8.4GHz/[O~IV], to identify the radio contribution from the Seyfert nuclei. We found that Seyfert 1 and Seyfert 2  galaxies are statistically similar in their radio emission, relative to the  strength of the AGN. On the other hand, when separated  by  the dominant source of [Ne~II] emission, we found that two groups, AGN and star formation dominated sources, are statistically different in their radio loudness.  We found that sources with strong star formation, relative to the strength of the AGN, or alternatively weak AGNs,  tend to be more radio loud, by a factor of $\sim 2.8$, than AGN dominated sources. Furthermore, in the 8.4GHz-[O~IV] correlation,  sources in the star formation dominated region have lower mass accretion rates than the rest of the sample. This result is in agreement with the anticorrelation found by  \citet{2002ApJ...564..120H} between the radio loudness and Eddington accretion rates. Given the sub-Eddington nature of the more radio loud sources in our sample, sources in the star formation dominated region, one  can invoke advection-dominated accretion  models  to explain their bright compact radio emission and a low-luminous nucleus both consistent with their high radio-to-mid-infrared parameter.

 We also found that high resolution 1.4~GHz emission could also be used as a reasonable proxy for the AGN radio contribution. When using the 1.4GHz/[O~IV] ratios we found that sources in which the [Ne~II] emission is dominated by the stellar activity, or alternatively weak AGNs,  tend to be more radio loud, by a factor of $\sim 3.0$, than AGN dominated sources, in close agreement with the radio loudness characterization using the radio core power at 8.4GHz. When comparing the observed 8.4~GHz and 1.4~GHz high-resolution core emission in our sample of Seyfert galaxies we found a strong  linear correlation, suggesting that the core radio emission is dominated by the active nuclei with a spectral index of $f_\nu \propto \nu^{-0.7}$. We found  the spectral index distribution, between 1.4~GHz and 8.4~GHz, for Seyfert 1 and Seyfert 2 galaxies and for AGN and star formation dominated sources to be statistically similar within the groups.

We suggest that the radio-to-optical parameter, $R$, used to classify  galaxy radio loudness may erroneously quantize  the radio contribution in galaxies with strong star formation or alternatively, weak AGNs, in which the optical luminosity may be contaminated by star formation and/or suffer from dust extinction. Instead, we found that a radio-to-mid-infrared parameter, 8.4GHz/[O~IV], is a better probe for the radio loudness of the galaxy,  as it is basically  unaffected by star formation contamination and is less affected by extinction than optically dependent  diagnostics.

\section*{Acknowledgments}
We thank the referee for a very useful comments that improved
the paper. We would also like to acknowledge  D.M. Crenshaw for valuable comments on the work. This research has made use of NASA's Astrophysics Data System. Also, this research has made use of the NASA/IPAC Extragalactic Database (NED) which is operated by the Jet Propulsion Laboratory, 
California Institute of Technology, under contract with the National Aeronautics and Space Administration. The VLA is operated by the US National Radio Astronomy Observatory which is operated by Associated
Universities, Inc., under cooperative agreement with the National Science Foundation. SMART was developed by the IRS Team at Cornell University and is available through the Spitzer Science Center at Caltech. Basic research in astronomy at the NRL is supported by 6.1 base funding.

\bibliographystyle{mn2e}

\bibliography{ms}

\appendix

\section[]{Spitzer/IRS Low-Resolution 5-35~$\mu$m Spectra of the Sample}
\begin{figure*}
\vbox to 174mm{\vfil  \includegraphics[width=174mm]{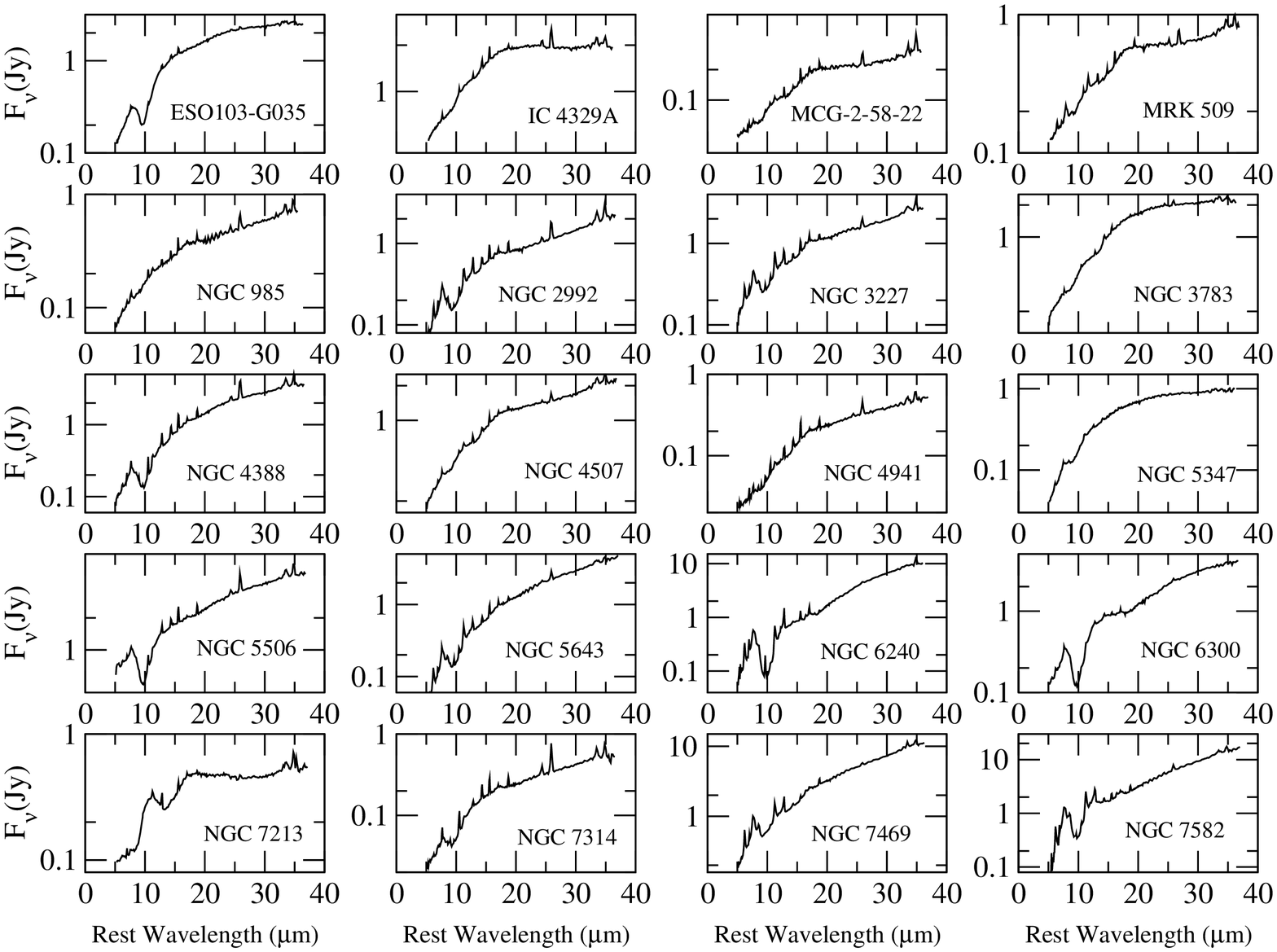}
    \caption{Spitzer/IRS low-resolution 5-35~$\mu$m spectra from our analysis of 20 sources  presented in \citet{2008ApJ...689...95M} and discussed in the present work\label{spectra}}
\vfil}
\end{figure*}

 The full sample presented in this work  includes  mid-infrared fluxes  from \citet{2007ApJ...671..124D}, \citet{2007arXiv0710.4448T}, \citet{2002A&A...393..821S}, \citet{2005ApJ...633..706W} and from our analysis of archival spectra observed with the Infrared Spectrograph (IRS) on board {\it Spitzer Space Telescope} \citep{2004ApJS..154....1W}.  Figure~\ref{spectra} shows the resulting IRS low-resolution spectra (Short-Low, R$\sim$60$-$127 and Long-Low, R$\sim$57$-$126) from our analysis of archival spectra for 20 sources presented in the present work. The mid- and far-infrared properties of this sample are discussed in detail in \citet{2008ApJ...689...95M}. For the   analysis of the mid-infrared spectra  observed with IRS/{\it Spitzer} we followed  the procedure described in \citet{2008ApJ...682...94M}.

\label{lastpage}

\end{document}